\documentclass[useAMS,usenatbib]{mn2e}

\usepackage {graphicx}
\def\aj{\,{AJ}}
\def\apj{\,{\rm ApJ}}

\def\apjs{\,{\rm ApJS}}

\def\mnras{\,{\rm MNRAS}}

\title[
Do star formation rates of galaxy clusters depend on mass?
]
{
Do star formation rates of galaxy clusters depend on mass?: blue/late-type fractions, total star formation rates of 115 galaxy clusters as a function of cluster virial mass
}
\author[T. Goto]
{Tomotsugu Goto$^{1}$\thanks{E-mail:tomo@jhu.edu}
  \\
  $^{1}$ Department of Physics and Astronomy, The Johns Hopkins
  University, 3400 North Charles Street, Baltimore, MD 21218-2686, USA
}

\begin{document}


\pagerange{\pageref{firstpage}--\pageref{lastpage}} \pubyear{2003}

\maketitle

\label{firstpage}

\begin{abstract}
 There has been plenty of observational evidence of cluster galaxy evolution such as the Butcher-Oemler effect and the decrease in S0 fraction with increasing redshift. As a modern version, the redshift evolution of total star formation rate (SFR) in clusters has been actively debated recently. However, these studies of redshift trend have been always hampered by the possible mass dependence; due to the observational selection effects, clusters found at higher redshift inevitably tend to be more massive and luminous than local counterparts. Consequently, one has to correct for the mass trend before interpreting the redshift trend. 
 As an attempt to break this degeneracy, we investigate the mass dependence of blue/late-type fractions and total SFR of 115 clusters at $z\leq 0.09$ selected from the Sloan Digital Sky Survey. 
  We find that none of blue/late-type fractions, total SFR and total SFR normalized by cluster mass shows significant dependence on cluster virial mass. The scatter is much larger at each cluster mass than a possible trend. Our results indicates that physical mechanisms that depend on cluster mass (such as the ram-pressure stripping) are not likely to be solely responsible for cluster galaxy evolution. Our results also provide an excellent low redshift comparison sample for future high redshift cluster SFR studies.

\end{abstract}


\section{Introduction}\label{intro}

 The Butcher-Oemler effect was first reported by Butcher and Oemler
 (1978, 1984) as an increase in the fraction of blue galaxies ($f_{\rm
 b}$) toward higher redshift in 33 galaxy clusters over the redshift range
 0$<z<$0.54.
 Butcher and Oemler's work had a strong impact since it showed direct evidence for the
 evolution of cluster galaxies. Later, Rakos and Schombert (1995) found that the
 fraction of blue galaxies increases from 20\% at $z$=0.4 to 80\% at
 $z$=0.9, suggesting that the evolution in clusters is even stronger than
 the original Butcher-Oemler effect. Margoniner and de Carvalho (2000) studied 48
 clusters in the redshift range of 0.03$<z<$0.38, and detected a strong
 Butcher-Oemler effect consistent with  that of Rakos and Schombert (1995). 

 Although the detection of the Butcher--Oemler effect has been claimed in
 various studies, 
 there have been some suggestions of strong selection biases in the cluster samples.
 Newberry, Kirshner, and Boroson (1988) measured the velocity
 dispersions and surface densities of galaxies in clusters, and found a
 marked difference between local clusters and intermediate 
 redshift clusters. More recently, Andreon and Ettori (1999)
 measured the X-ray surface brightness profiles, sizes and luminosities of the
 Butcher--Oemler sample of clusters, and concluded that the sample is not
 uniform. 
      It is a concern that these selection biases may smear evolutionary trends. Especially, Margoniner et al. (2001) and Goto et al. (2003a) found that the blue fraction depends on cluster richness. If richer, and thus more massive, clusters have lower fractions of blue galaxies, cluster samples with a Malmquist-type bias naturally have an erroneously weaker redshift trend since distant cluster tend to be more massive due to the selection effect.
 Thus, in order to reveal the true redshift evolution, it is important to clarify if blue fraction depends on cluster mass or not.

 Recently, as a modern version of the Butcher-Oemler effect, the evolution the star formation rate (SFR) in galaxy clusters has been actively debated (e.g., Kodama \& Bower 2001). 
     Postman et al. (1998;2001) found that a large number of cluster members show high levels of star formation activity, and that the average SFR is higher in $z\sim 0.9$ clusters than in low redshift clusters.
      Finn et al. (2004) imaged the galaxy cluster Cl0023+0423B at $z=0.845$ through narrow-band (H$\alpha$) filter and found that the integrated SFR normalized by cluster mass ($\Sigma SFR/M_{cl}$) is a factor of 10 higher than that of $z\sim0.2$ clusters in the literature. 
However, these studies have been hampered by the same degeneracy between mass dependence and redshift dependence; Finn et al. (2004) also found  $\Sigma SFR/M_{cl}$ depends on cluster mass, in addition to the redshift. 
 Therefore, it has not been clear if the variation in cluster SFR is caused by cluster mass or redshift evolution.


 In this Letter, we aim to clarify the mass dependence of  the Butcher--Oemler effect (blue fractions) and the total SFR of clusters using an extremely large number of clusters at low redshift.
 Recently, using 250,000 spectra in the Sloan Digital Sky Survey Data Relase 2 (SDSS; Abazajian et al. 2004), a large cluster catalog with 335 clusters with at least 20 spectroscopic members has been compiled (Goto 2005a; see also Goto et al. 2002a,b for a SDSS cluster catalog). This catalog provides us with a good opportunity to investigate whether the blue fraction and $\Sigma SFR$ depend on cluster mass or not using a statistically large sample (c.f., previous samples consisted of only a few to dozens of clusters). The results provide an important constraint on the physical mechanism associated with cluster mass,  in addition to an important basis in studying redshift evolution of these quantities using high redshift clusters.

%

 The cosmological parameters adopted throughout this paper are $H_0$=71 km
 s$^{-1}$ Mpc$^{-1}$ and ($\Omega_m$,$\Omega_{\Lambda}$)=(0.27,0.73) (Bennett et al. 2003).


\section{Data \& Analysis}\label{data}

\begin{figure}
\includegraphics[scale=0.6]{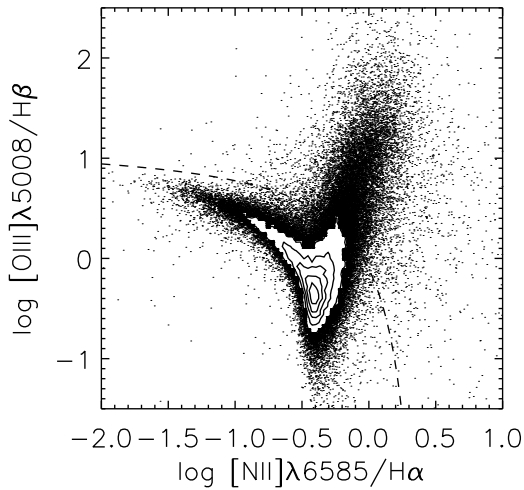}
\caption{ Emission line ratios used to remove AGNs from our sample. The dashed line is the criterion between starbursts and AGNs described in Kewley et al. (2001). Galaxies with line ratios higher than the dashed line are removed from our sample as AGNs.
}\label{fig:Osterbrook} 
\end{figure}

 Recently, a large cluster catalog of $\sim 335$ clusters with at least 20 spectroscopic members has been compiled by Goto (2005a) using $\sim$250,000 spectra of the SDSS DR2. From this catalog, we select clusters with $z\leq 0.09$ and $\geq$20 spectroscopic members. This redshift range allows us to probe member galaxies as faint as $M_r=-20.38$ in a volume limited way since the SDSS is flux-limited spectroscopic survey to the depth of $r=17.77$.   We use Petrosian magnitudes k-corrected using Blanton et al. (2003;v3\_2) and corrected for galactic extinction using reddening map of Schlegel, Finkbeiner \& Davis (1998). The actual number of clusters used is 115, which is the largest number of spectroscopic clusters used to study the Butcher-Oemler effect. We restrict cluster member galaxies to those within the virial radius of the cluster (computed using Girardi et al. 1998), and within 3$\sigma$ of the velocity dispersion. This radial normalization is important since the fraction of blue/late-type
 galaxies is a strong function of radius (e.g., Goto et al. 2003c; Goto et al. 2004), and thus, using different radius of each cluster could produce a false trend.
 Assuming each cluster is virialized, we compute virial mass ($M_V$) of each cluster using the virial radius and the velocity dispersion (given in Goto et al. 2005a) using the prescription given in Girardi et al. (1998). 

 We also compute SFR of each cluster galaxies from H$\alpha$ luminosity (measured using  Goto et al. 2003b) using the prescription given in Hopkins et al. (2003). Goto et al. (2005a) measured H$\alpha$ flux using the flux summing technique described in Goto et al. (2003b). In order to correct for stellar absorption, we used H$\delta$ equivalent width (EW) measured in Goto et al. (2005a), and converted it to the stellar absorption at H$\alpha$ and H$\beta$ using the following empirical relation (Miller \& Owen 2002): H$\beta$ EW$_{absorption}$  is equal to H$\delta$ EW$_{absorption}$ and H$\alpha$ EW$_{absorption}$ is equal to 1.3 + $0.4\times$  H$\delta$ EW$_{absorption}$ (Keel 1983).  Using the stellar absorption corrected H$\alpha$/H$\beta$ flux ratio, we correct for dust extinction using the prescription given in Hopkins et al. (2003). We apply this extinction correction only when both  H$\alpha$ and H$\beta$ lines are in emission. Finally we apply the 3" fiber aperture correction by scaling the H$\alpha$ flux by the ratio of the Petrosian $r$-band flux to that measured within the 3'' fiber. 
 By using this stellar absorption, dust extinction, aperture corrected H$\alpha$ flux to the equation given in Kennicutt (1998), we obtain a SFR for each cluster galaxy.

 When we compute total SFRs, we exclude AGNs using the prescription given in Kewley et al. (2001). Fig.\ref{fig:Osterbrook} shows line ratios of [OIII](5008\AA)/H$\beta$ versus [NII](6585\AA)/H$\alpha$. The dotted line is the criterion used to divide AGNs and starbursts proposed by Kewley et al. (2001). Here, we use the flux measured by the SDSS pipeline (Stoughton et al. 2002) since for strong emission lines, the Gaussian fit by the pipeline works well (Goto et al. 2003b).  We take a conservative approach in excluding AGNs. We only exclude a galaxy as an AGN when all four lines are detected in emission and the line ratio is above the dotted line in Fig.\ref{fig:Osterbrook}. We have also checked that rejecting more AGNs using only 2 lines does not change our results.

\section{Results}\label{results}

\subsection{Blue/Late-type Fraction as a Function of Mass}

\begin{figure}
\includegraphics[scale=0.6]{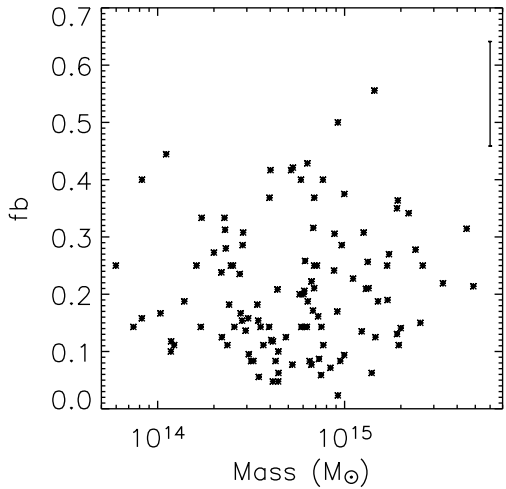}
\caption{Fraction of blue galaxies as a function of cluster mass, A typical size of error bars is shown at the upper right corner of the plot.  Spearman's correlation coefficient is 0.09.
}\label{fig:fb} 
 \end{figure}

 In this section, we investigate if the blue fraction (the Butcher-Oemler effect) depends on cluster mass. We separate blue and red galaxies using the restframe $u-r=2.22$ colour. Strateva et al. (2001) showed that this colour is the local minimum of the bimodal distribution of red/blue galaxies (Baldry et al. 2004), and thus it separates early- and late-type galaxies well at $z<0.4$.
 Fig. \ref{fig:fb} shows the ratio of blue ($u-r<2.22$) galaxies to all galaxies in each cluster as a function of cluster mass. A median size of the error bars is shown in the upper right corner of the plot. As can be seen, there is no obvious trend of the blue fraction with cluster mass. Indeed,   Spearman's correlation coefficient is as small as 0.09.
 
\begin{figure}
\includegraphics[scale=0.6]{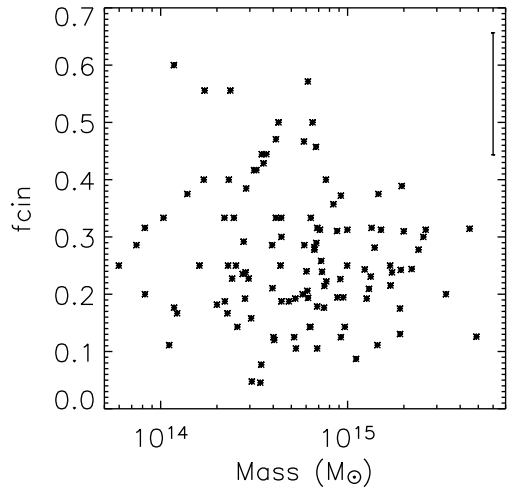}
\caption{Fraction of late-type galaxies ($Cin>0.4$) as a function of cluster mass. A typical size of error bars is shown at the upper right corner of the plot.  Spearman's correlation coefficient is -0.08.
}\label{fig:fcin} 
\end{figure}

 In Fig.\ref{fig:fcin}, we plot the fraction of morphologically late-type galaxies ($f_{Cin}$) as a function of cluster mass. Here, late-type galaxies are selected using the concentration parameter ($Cin$), which is defined as a ratio of  Petrosian 50\% light radius to Petrosian 90\% light radius. This parameter, $Cin$, is known to be well correlated with eye-classified morphology (Shimasaku et al. 2001; Strateva et al. 2001), and thus has been used in many previous studies (e.g., Gomez et al. 2003; Kauffman et al. 2004; Tanaka et al. 2004). We regard galaxies with $Cin>0.4$ as late-type galaxies and compare the fraction to the number of all galaxies.
 As in the previous figure, the median error bar is shown in the upper right corner of Fig.\ref{fig:fcin}. There is no clear trend of mass dependence in Fig.\ref{fig:fcin}.  Spearman's correlation coefficient is -0.08.

\subsection{Total SFR as a Function of Cluster Mass}

 Next, we investigate the SFR in clusters as a function of cluster mass. The SFRs of cluster member galaxies are computed using the method described in Section \ref{data}. 

\begin{figure}
\includegraphics[scale=0.6]{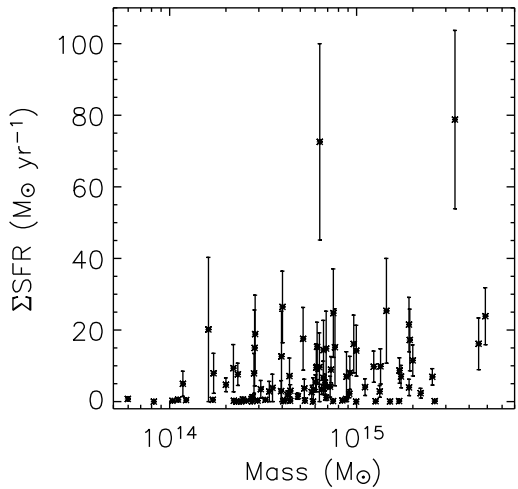}
\caption{Total SFR as a function of cluster mass. Errors are based on the number of spectroscopic galaxies with $M_r\leq-20.38$ in each cluster. Incompleteness in spectroscopic observation is corrected using the number ratio of galaxies with spectra and galaxies in the SDSS imaging data. Spearman's correlation coefficient is 0.23. 
}\label{fig:total_SFR} 
\end{figure}

In Fig.\ref{fig:total_SFR}, we sum all the SFRs of cluster member galaxies with $M_r\leq-20.38$ to plot total amount of SFRs in a  cluster against cluster mass. Here, we correct for incompleteness in spectroscopic targeting of the SDSS by multiplying by the ratio of all galaxies in angular area of the cluster in the imaging data  to all spectroscopic galaxies in the same region of the sky.
 Since we do not know the redshift for galaxies in the imaging data, we use magnitude range of $r<17.77$ in correcting spectroscopic incompleteness, i.e., we use all spectroscopic galaxies regardless of the redshift and all the galaxies in the imaging data with $r<17.77$. This correction is about factor of 2-3, and does not depend on cluster mass. 
 Due to this correction, although the SDSS does not take spectra of all galaxies in the cluster area (the adjacent SDSS fibers cannot be closer than 55''), the computed amount of total SFR is statistically corrected to the true total amount of SFR in cluster region in the range of $M_r\leq-20.38$. The error bars shown in the figure are based on the number of galaxies contributing to the sum.
  In Fig.\ref{fig:total_SFR}, there is no apparent correlation between total SFR ($\Sigma SFR$) and the cluster mass. Spearman's correlation coefficient is 0.23. 

\begin{figure}
\includegraphics[scale=0.6]{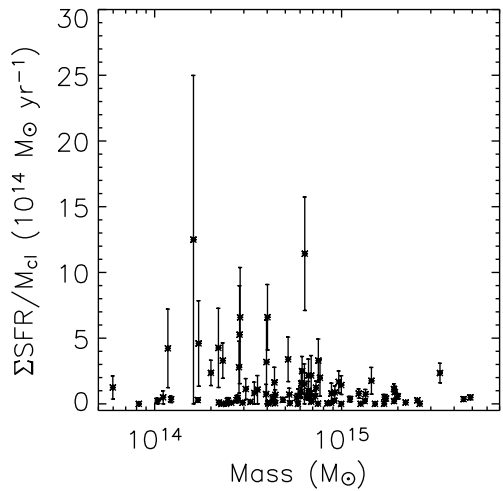}
\caption{Total SFR normalized by cluster mass. Errors are based on the number of spectroscopic galaxies in each cluster. Incompleteness in spectroscopic observation is corrected using the number ratio of galaxies with spectra and galaxies in the SDSS imaging data. Spearman's correlation coefficient is  -0.12.
}\label{fig:SFR/mass} 
\end{figure}

 Following a recent trend (Finn et al. 2004; Kodama et al. 2004), we plot $\Sigma SFR$ normalized by cluster mass ($\Sigma SFR/M_{cl}$) in Fig.\ref{fig:SFR/mass}. The plot looks as if there is a decreasing trend in $\Sigma SFR/M_{cl}$ with increasing cluster mass. However,  Spearman's correlation coefficient is as small as  -0.12, and thus, the apparent correlation is hardly significant. A more notable feature is the large amount of scatter among the cluster sample.

\section{Discussion}

 In Section \ref{results}, we found that both blue fractions and late-type fractions do not significantly depend on cluster mass (Figs. \ref{fig:fb} and \ref{fig:fcin}). We also found that $\Sigma SFR$ and $\Sigma SFR/M_{cl}$ do not depend on cluster mass (Figs \ref{fig:total_SFR} and \ref{fig:SFR/mass}). 
   It has been known that if blue fractions or SFR of clusters depend on cluster mass, such observational evidence could be interpreted as evidence for the ram-pressure stripping mechanism (Fujita 2004; Fujita \& Goto 2004), which mainly alters the SFR of galaxies, and thus, the blue fraction. In a simple estimation, the ram-pressure is proportional to $\rho v^2$. $L_{\rm X}$ is proportional to $\rho^2$. From the virial theorem, $v^2\propto T$. The $L_{\rm X}$--$T$ relation studied by Xue and Wu (2000) is $L_{\rm X}\propto T^{2.8}$.
 Therefore, the ram-pressure is proportional to $\sim L_{\rm X}^{0.86}$, i.e., the ram-pressure is stronger for more X-ray luminous, and thus massive, clusters. Therefore, if the ram-pressure stripping is the only mechanism responsible for the colour evolution of galaxies in clusters, the 
 fraction of blue galaxies should always be lower in more X-ray luminous 
 clusters,  which therefore are more massive clusters (Fujita \& Nagashima 1999). Therefore, our findings of no mass dependence of blue fractions and $\Sigma SFR$ suggests that the ram-pressure stripping is not the only dominant physical mechanism that drives the cluster galaxy colour evolution.

 In addition, tidal stripping by the gravitational potential of the cluster can also strip gas in the galaxy disk, and consequently can halt the SFR in the galaxy (e.g., Gnedin 2003a,b).  This tidal stripping effect (again mainly affects blue fractions and $\Sigma SFR$) should also depend on the cluster mass.  Statistically, stronger tidal forces are expected in more massive clusters, and thus, massive clusters should have more evolved galaxies (see equation 16 in Gnedin 2003a).  In fact, in the simulation by Moore et al. (1999), stronger tidal evolution is found in larger clusters. Therefore,  our findings may indicate that tidal stripping may not be a key physical mechanism.
  Similarly, galaxy-galaxy merging should also be more effective in clusters with smaller velocity dispersion (Makino \& Hut 1997); and thus, the galaxy-galaxy merging  mechanism, which affects galaxy morphology mainly, may be dis-favored as well by our discovery of no mass dependence of the late-type fractions.

 Our results, however, are inconsistent with some previous work which found that blue fractions are lower for richer clusters (Margoniner et al. 2001; Goto et al. 2003a). 
  Possible sources of discrepancy are: (i) their use of imaging data with a background subtraction technique in assessing the blue fractions; (ii) different magnitude range probed, i.e., Margoniner et al. (2001) probed as faint as $M^* + 2$, whereas our work is limited to $M_r\leq-20.38$. Especially, Tanaka et al. (2005) found different morphology-SFR-density relations for faint and bright galaxies, and thus, the difference in the magnitude range is a possible source of the apparent discrepancy.
 Mart{\'{\i}}nez et al. (2002) found a strong correlation between the relative fraction of galaxies with high SFR and the parent group mass using 2209 galaxy groups found in 2dF. However, their sample consisted of poor groups with not much overlap with the mass range we probed. 
 On the other hand,  Fairley et al. (2002) studied eight X-ray selected
 clusters, and did not find any dependence of blue fractions on X-ray
 luminosities. Recently, De Propris et al. (2004) found that blue fractions in 60 clusters in the 2dFGRS do not depend on cluster velocity dispersions. These results are consistent with ours although they are based on fewer number of clusters.

 Our results imply that there is no strong need to correct for the mass dependence when one studies redshift evolution of $f_b$ or $\Sigma SFR$. However, the large scatter seen in Figs. \ref{fig:fb}-\ref{fig:SFR/mass} cautions not to mis-interpret a noise caused by the scatter as a redshift trend. The sample presented in Finn et al. (2004) and Kodama et al. (2004) consists of only 4 and 5 clusters. Eagerly awaited are statistical numbers of clusters at high redshift with SFR measurements.

 Finally this work provides a good local comparison sample to future high redshift cluster studies. Since currently available high-z cluster $\Sigma SFR$s are limited by the H$\alpha$ luminosity (due to the narrow band H$\alpha$ imaging with no strict absolute magnitude limit), it is not straight-forward to compare. However, with all the caveats in mind, if we forcefully apply suggested corrections of 70-80\% in  $\Sigma SFR$ between a H$\alpha$ imaging survey and a spectroscopic survey (Homeier et al. in prep.), majority of our clusters has estimated $\Sigma SFR/M_{cl}$ of $<20$ $\times 10^{14}$ M$_{\odot}$, which is much smaller than that of $z=0.8$ (Finn et al. 2004). However, a truly meaningful comparison has to be postponed until one has both low and high redshift sample observed with the same technique. Once one has the $\Sigma SFR$ of high redshift clusters applying the magnitude limit of $M_r\leq-20.38$, it should be easy to compare with our low redshift sample, and to study redshift
 evolution.



\section{Conclusions}
 
 We have investigated blue/late-type fractions, total SFR and total SFR normalized by mass for 115 clusters ($z\leq 0.09$) carefully selected from the SDSS DR2. We found that all of the above quantities do not significantly depend on cluster virial mass. Therefore, our results suggests that the physical mechanisms that depend on cluster mass are not favored to be responsible for cluster galaxy evolution. The lack of mass trend implies that previously found redshift trends can be interpreted as a redshift evolution instead of mass dependence. However, our findings of large scatter in each quantity cautions us not to mis-interpret noise caused by the scatter as a redshift trend. In order to investigate redshift evolution of cluster SFRs, much larger samples of high redshift clusters are needed. Our results provide an excellent benchmark for such future high redshift cluster studies.

\section*{Acknowledgments}

We thank Dr. Ani Thakar for his friendly help in downloading the publicly available SDSS data.
 We are grateful to Oleg Gnedin for useful discussion.
Funding for the creation and distribution of the SDSS Archive has been provided by the Alfred P. Sloan Foundation, the Participating Institutions, the National Aeronautics and Space Administration, the National Science Foundation, the U.S. Department of Energy, the Japanese Monbukagakusho, and the Max Planck Society. The SDSS Web site is http://www.sdss.org/. 

The SDSS is managed by the Astrophysical Research Consortium (ARC) for the Participating Institutions. The Participating Institutions are The University of Chicago, Fermilab, the Institute for Advanced Study, the Japan Participation Group, The Johns Hopkins University, Korean Scientist Group, Los Alamos National Laboratory, the Max-Planck-Institute for Astronomy (MPIA), the Max-Planck-Institute for Astrophysics (MPA), New Mexico State University, University of Pittsburgh, Princeton University, the United States Naval Observatory, and the University of Washington.


\label{lastpage}

\end{document}